\documentclass[conference,twoside]{IEEEtran} 

\usepackage[utf8]{inputenc}
\usepackage{ amsmath, graphicx, amssymb, amsthm, mathtools, units, algorithmic}
\usepackage[boxed]{algorithm}
\usepackage[english]{babel}
\usepackage{url,cite}

\usepackage[caption=false,font=scriptsize]{subfig}

\usepackage{tikz,pgfplots}
\usetikzlibrary{arrows,shapes.geometric,calc,decorations.markings,patterns}
\pgfplotsset{compat=newest}

\definecolor{mathblue}{HTML}{3F3D9A}
\definecolor{mathpurp}{HTML}{9A3D71}
\definecolor{mathsand}{HTML}{9A8C3D}
\definecolor{mathgrn}{HTML}{0F9F4F}

\def\bs{\boldsymbol}
\def\cl{\mathcal}
\def\bb{\mathbb}
\def\ts{\textstyle}
\def\R{{\bb R}}
\def\A{\bs A}

\def\y{\bs y}
\def\x{\bs x}
\def\xx{\bs \xi}
\def\ie{{\em i.e.},~}
\def\eg{{\em e.g.},~}
\def\etal{{\em et al.}~}
\def\g{\bs g}
\def\gg{{\bs \gamma}}
\def\z{\bs z}
\def\Z{\bs Z}
\def\Ponep{\bs P_{\vone^\perp_m}}

\def\I{\bs I}
\def\vone{{\bs 1}}

\def\v{\bs v}
\def\diag{{\rm diag}}
\def\argmin{{\mathop{\rm argmin}}}
\graphicspath{{./fig/}}

\floatname{algorithm}{Algorithm}

\title{A Greedy Blind Calibration Method for Compressed Sensing with Unknown Sensor Gains} 
\IEEEoverridecommandlockouts
\author{\IEEEauthorblockN{Valerio Cambareri, Amirafshar Moshtaghpour, Laurent Jacques}
	\IEEEauthorblockA{Image and Signal Processing Group, ICTEAM/ELEN,\\
		Universit\'e catholique de Louvain,	Louvain-la-Neuve, Belgium.\\
		E-mail: {\{valerio.cambareri,~amirafshar.moshtaghpour,~laurent.jacques\}@uclouvain.be}
		\thanks{The authors are funded by the Belgian F.R.S.-FNRS. Part of this study is funded by the project {\sc AlterSense} (MIS-FNRS).}
	}
}

\begin{document}
	\maketitle
	\begin{abstract}
		The realisation of sensing modalities based on the principles of compressed sensing is often hindered by discrepancies between the mathematical model of its sensing operator, which is necessary during signal recovery, and its actual physical implementation, which can amply differ from the assumed model. In this paper we tackle the bilinear inverse problem of recovering a sparse input signal and some unknown, unstructured multiplicative factors affecting the sensors that capture each compressive measurement. Our methodology relies on collecting a few snapshots under new draws of the sensing operator, and applying a greedy algorithm based on projected gradient descent and the principles of iterative hard thresholding. We explore empirically the sample complexity requirements of this algorithm by testing its phase transition, and show in a practically relevant instance of this problem for compressive imaging that the exact solution can be obtained with only a few snapshots.
	\end{abstract}
	
	\section{Introduction}
	\label{sec:intro}
	The implementation of practical sensing schemes based on Compressed Sensing (CS) \cite{Donoho2006} often encounters physical non-idealities in realising the mathematical model of the sensing operator, whose accuracy is paramount to attaining a high-quality recovery of the observed signal \cite{HermanStrohmer2010}. Among such non-idealities, we here focus on the case in which each compressive measurement is affected by an unknown multiplicative factor or {\em sensor gain}, \ie we focus on the sensing model
	\begin{equation} 
	\label{eq:measurement-model-matrix}
	\bs y_{l} = {\rm diag}(\bs g) \bs A_{l} \bs x, \ l \in [p] \coloneqq \{ 1, \ldots, p\},
	\end{equation}  
	where $\x \in \R^n$ is the input signal, $\A_l \in \R^{m\times n}$, $l \in [p]$ are independent and identically distributed (i.i.d.)~random sensing matrices, and $\bs y_l \in \R^m$, $l \in [p]$ are the respective {\em snapshots} of measurements obtained by applying each sensing matrix to $\x$ (the reason why the acquisition is partitioned in snapshots will be cleared below). 
	In this {\em uncalibrated} sensing model $\g \in \R^m_+$ is an unknown set of positive-valued gains that remains identical throughout the snapshots, but whose value is unknown. 
	Hence, this sensing model is {\em bilinear} in $\x$ and $\g$, and retrieving both quantities given the measurements is a non-trivial~\emph{bilinear inverse problem} (BIP). Note that \eqref{eq:measurement-model-matrix} can be practically realised in compressive imaging schemes using snapshot (\ie parallel) acquisition by convolving an input signal with one or more random masks, such as those detailed in \cite{Romberg2009,BjorklundMagli2013,LiutkusMartinaPopoffEtAl2014,BahmaniRomberg2015,DumasLodhiBajwaEtAl2016}. When sensor gains are not calibrated,~\eg in the presence of fixed-pattern noise or strong pixel-response non-uniformity \cite{HayatTorresArmstrongEtAl1999}, taking a few snapshots allows for on-line blind calibration without missing any instance of the signal $\x$ due to an off-line calibration process, as we showed in previous contributions \cite{CambareriJacques2016,CambareriJacques2016a}. There, we proved that instances of~\eqref{eq:measurement-model-matrix} with sensing matrices having i.i.d.~sub-Gaussian entries (for a rigorous definition, see \cite{Vershynin2012}) and $(\x,\g)$ being either unstructured or endowed with subspace models can be solved by a simple, suitably initialised projected gradient descent on a non-convex objective. The number of measurements ensuring the recovery of the exact solution was shown to be\footnote{Hereafter, given two functions $f,g$, $f\gtrsim g$ indicates that $f > C g$ for some constant $C>0$.} $m p \gtrsim n+m$, \ie a linear {\em sample complexity} in the dimensions of the unknowns (up to $\log$ factors, and referring to the findings in \cite{CambareriJacques2016a}). 
	
	In this paper we focus on the case in which the single input signal $\x$ has a $k$-sparse representation in a known basis. To leverage this more involved model on $\x$ we simply resort to a hard thresholding operator at each iterate of our former non-convex algorithm, turning it into a greedy scheme. 
	The proposed greedy approach allows for blind calibration in actual CS schemes; the additional requirement of our methodology is a set of $p$ snapshots that collects a sufficient amount of information on $(\x, \g)$. 
	Our emphasis is on assessing, at least empirically, how the sample complexity can be reduced in function of the signal-domain sparsity $k$ (up to $\log$ factors). Hence, provided $\x$ is sufficiently sparse, we will show empirically that the total amount of measurements $m p$ can be lower than $n$ while still recovering both $(\x,\g)$. 
	
	\subsection{Related Work}
	Blind calibration of sensor gains has been tackled in recent literature, starting from initial approaches for uncalibrated sensor networks in \cite{BalzanoNowak2008,LiporBalzano2014}, and more recently for radio-interferometry~\cite{RepettiBirdiDabbechEtAl2017}. In the context of CS, some algorithms have been proposed to cope with such model errors \cite{ZhuLeusGiannakis2011,ParkerCevherSchniter2011,BilenPuyGribonvalEtAl2014,FriedlanderStrohmer2014,SchuelkeCaltagironeZdeborova2015}. Interestingly, most algorithms use sparse or known subspace models for \emph{several} input signals, rather than random draws of the sensing operator itself (as typically feasible in optical CS schemes~\cite{BahmaniRomberg2015,DumasLodhiBajwaEtAl2016}); moreover, these works do not attain sample complexity results that grant exact recovery. A first work proposing such provable guarantees using a single sparse input signal was introduced by Ling and Strohmer \cite{LingStrohmer2015} based on a {\em lifting} approach to the problem (as in \cite{AhmedRechtRomberg2014}; improved guarantees were outlined in~\cite{StoegerJungKrahmer2016}). The main drawback of this approach is its computational complexity, given that it corresponds to very large-scale semidefinite programming. 
	
	Our former contributions \cite{CambareriJacques2016,CambareriJacques2016a} then showed that a non-convex approach could provide exact recovery guarantees and computational advantages with respect to (w.r.t.) lifting approaches; these were inspired by the methodologies of Cand\`es \etal \cite{CandesLiSoltanolkotabi2015}, Sanghavi \etal \cite{SanghaviWardWhite2016}, and Sun~\etal\cite{SunQuWright2016} used for the closely related problem of phase retrieval. 
	
	For what concerns the related task of blind deconvolution, very recent approaches to this BIP adopt similar non-convex schemes~\cite{LiLingStrohmerEtAl2016,MansourKamilov2016} or alternating minimisation \cite{LeeTianRomberg2016,LeeLiJungeEtAl2017}, yet targeting a more general context than blind calibration and therefore subject to different requirements and conditions than those we encountered independently. Ling and Strohmer~\cite{LingStrohmer2016}  proposed linear least squares for settings including~\eqref{eq:measurement-model-matrix}.
	
	What is not covered, as we study a practical, non-convex solver for blind calibration under the sensing model~\eqref{eq:measurement-model-matrix} and sparse signal priors, is the {\em identifiability} of our BIP, \ie to what extent the solution $(\bs x,\bs g)$  can be \emph{uniquely and unambiguously} determined given $\bs y$; for completeness, we refer the reader to recent contributions on this aspect~\cite{ChoudharyMitra2013,ChoudharyMitra2014,LiLeeBresler2016,LiLeeBresler2017}.
	\subsection{Contributions and Outline}
	Our paper extends the non-convex algorithm devised in \cite{CambareriJacques2016,CambareriJacques2016a} to account for a sparse model in the signal domain; this is a fundamental prior for CS, whereas sparse models on the gains $\g$ could be inapplicable when these are drawn at random as each sensor is manufactured. Thus, we adopt a greedy algorithm to enforce signal-domain sparsity, and detail its empirical performances as a function of our BIP's dimensions. 
	
	Our findings are presented as follows. In Sec.~\ref{sec:ncbc} we introduce the non-convex problem and propose a greedy algorithm based on hard thresholding. This algorithm is studied numerically in Sec.~\ref{sec:pht}, where we focus on the empirical phase transition as the problem dimensions vary. We then simulate a practical case of blind calibration for compressive imaging in Sec.~\ref{sec:img}. A conclusion is drawn afterwards.
	
	\section{A Greedy and Non-Convex Approach \newline to Blind Calibration}
	\label{sec:ncbc}
	Our initial approach to the blind calibration problem involved defining a simple Euclidean loss, 
	\begin{equation}
	f(\xx,\gg) \coloneqq  \tfrac{1}{2 m p} {\ts \sum_{l=1}^{p} \left\Vert \diag(\gg) \A_l \xx - \y_l\right\Vert^2_2},
	\end{equation}
	and solving 
	\begin{equation}%
	\label{eq:bcpinit}
	(\hat{\x}, \hat{\g}) = {\mathop{\argmin}_{\xx \in \R^n, \gg \in \Pi_+^m}} f(\xx,\gg)
	\end{equation}%
	where $\Pi^m_+ \coloneqq \{\v \in \R^m_+ : \vone^\top_m \v = m\}$ is the {\em scaled probability simplex} and $\vone_m$ the vector of ones in $\R^m$. To begin with, up to a scaling all points in 
	\begin{equation}
	\big\{(\xx, \gg) \in \R^n \times \R^m : \xx = \alpha \x, \gg = \tfrac{\g}{\alpha}, \alpha \neq 0\big\}
	\end{equation}
	are minimisers of $f(\xx,\gg)$ (\ie the scaling of $(\x,\g)$ is anyway unrecoverable), so we adopted the constraint $\gg \in \Pi_+^m$ which fixes one admitted solution for $\alpha = \tfrac{\|\g\|_1}{m}$. This also serves to control $\|\gg\|_1$ during the iterates of our algorithm. 
	We then assume that $\x$ is $k$-sparse~w.r.t.~an orthonormal basis $\Z \in \R^{n\times n}$ such that $\x = \Z \z \in \Z \Sigma^n_k$, with $\Sigma^{n}_k \coloneqq \{\bs u \in \R^n : k = |{\rm supp} \, \bs u| \}$. 
	Thus, to enforce sparsity we aim to solve
	\begin{equation}%
	\label{eq:bcp}
	(\hat{\x}, \hat{\g}) = {\mathop{\argmin}_{\xx \in \Z \Sigma^n_k, \gg \in \Pi_+^m}} f(\xx,\gg),
	\end{equation}%
	where the problem would be non-convex both due to the bilinear nature of $f(\xx,\gg)$ and to that of the \emph{union} of $k$-dimensional canonical subspaces $\Sigma^n_k$. This differs w.r.t. the solver to \eqref{eq:bcpinit} we devised in \cite{CambareriJacques2016a}, as there we assumed the support $T \coloneqq {\rm supp} \ \bs z$ was given, \ie a \emph{known subspace model}. 
	
	We now proceed to devise an algorithm solving~\eqref{eq:bcp} that accounts for the two constraints. %
	\begin{algorithm}[!t]
		\caption{\label{alg1} Blind Calibration with Iterative Hard Thresholding (BC-IHT)\vspace{-4mm}}
		\begin{algorithmic}[1]
			\STATE Initialise $\xx_0 \coloneqq \tfrac{1}{m p}
			\sum^{p}_{l = 1} \left(\A_l\right)^\top \y_l$; $\gg_0 \coloneqq \vone_m$; the exact sparsity level $k$;  $j \coloneqq 0$. 
			\WHILE{stop criteria not met}
			\STATE \COMMENT{Line searches} \\ $\mu_{\xx} \coloneqq \argmin_{\upsilon \in \R} f(\xx_{j}-\upsilon\bs\nabla_{\xx} f({\xx_{j}, \gg_{j}}), \gg_{j})$ \\
			$\mu_{\gg} \coloneqq \argmin_{\upsilon \in \R} f(\xx_{j},\gg_{j}-\upsilon\bs\nabla^\perp_{\gg} f({\xx_{j}, \gg_{j}}))$ 
			\STATE \COMMENT{Hard-thresholded signal estimate} \\
			$\xx_{j+1} \coloneqq \Z {\cl H}_k\big[\Z^\top({\xx}_{j} - \mu_{\xx} \bs\nabla_{\xx} f({\xx_{j}, \gg_{j}}))\big]$
			\STATE \COMMENT{Gain estimate} \\
			${\gg}_{j+1} \coloneqq {\cl P}_{\cl G_\rho}\big[\gg_{j} - \mu_{\gg} \, \bs\nabla^\perp_{\gg} f({\xx_{j}, \gg_{j}})\big]$ 
			\STATE $j \coloneqq j + 1$
			\ENDWHILE
		\end{algorithmic}
	\end{algorithm}
	\begin{figure*}[!t]
		\null\hfill
		\subfloat[$n=2^9$, $k = 2^5$]{		
			\includegraphics{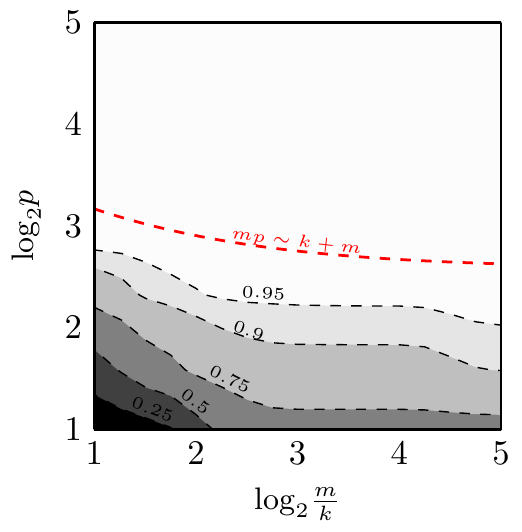}
		}\hfill
		\subfloat[$n=2^9$, $k = 2^6$]{	
			\includegraphics{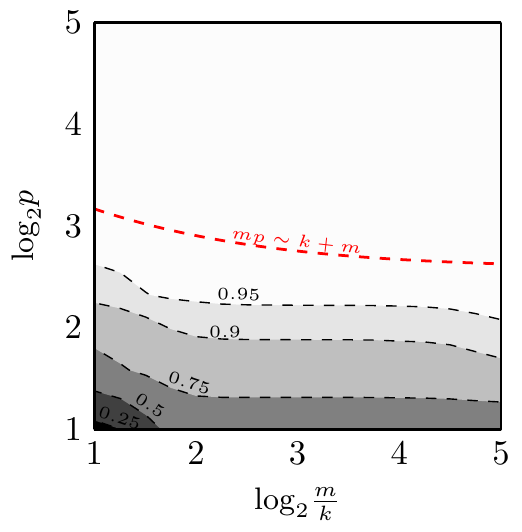}
		}\hfill
		\subfloat[$n=2^9$, $k = 2^7$]{
			\includegraphics{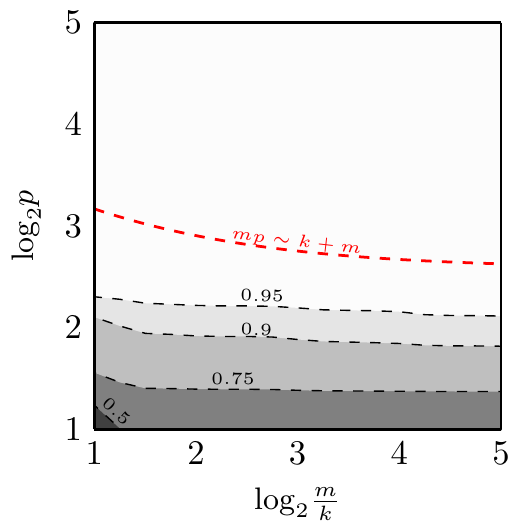}
		}\hfill
		\null\\
		\null\hfill
		\subfloat[$n=2^{10}$, $k = 2^5$]{
			\includegraphics{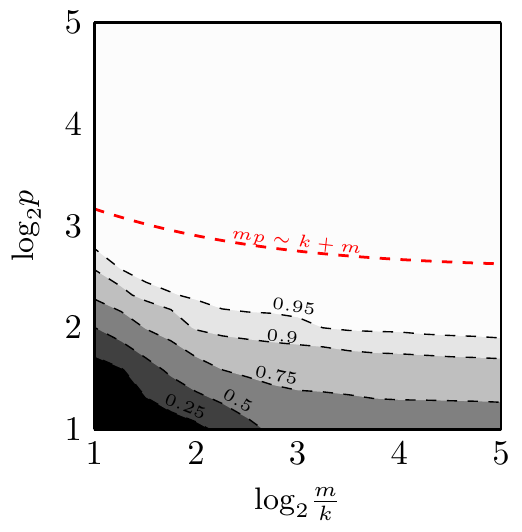}
		}\hfill
		\subfloat[$n=2^{10}$, $k = 2^6$]{
			\includegraphics{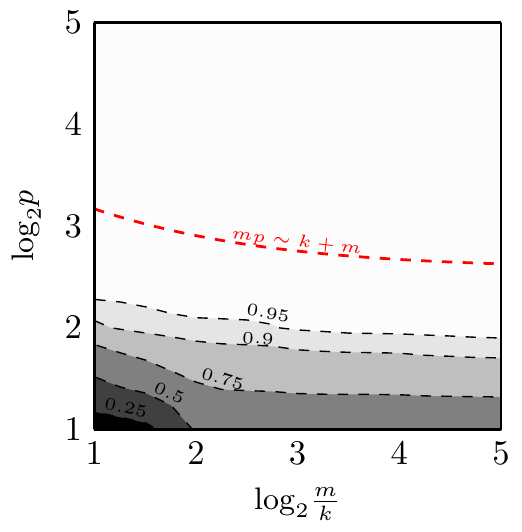}
		}\hfill
		\subfloat[$n=2^{10}$, $k = 2^7$]{
			\includegraphics{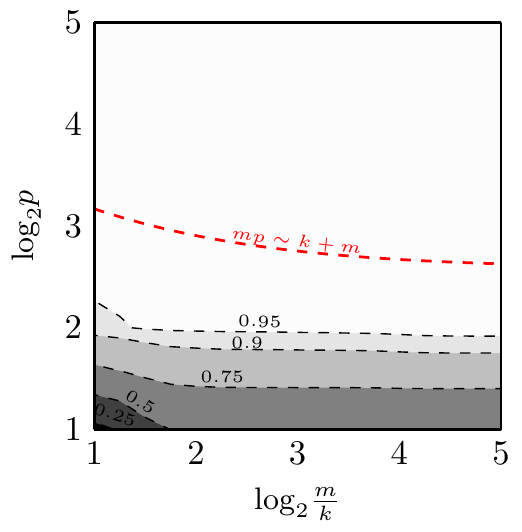}
		}\hfill
		\null
		\caption{\label{fig:ept}Empirical phase transition of Alg.~\ref{alg1} as $n$ increases (top to bottom) and $k$ increases (left to right), as a function of $\tfrac{m}{k},p$ and fixing $\rho = \tfrac{1}{2}$. We report the estimated contour levels of the probability of successful recovery, as it exceeds the value indicated above of each curve.\vspace{-1.5ex}}
	\end{figure*}
	%
	Firstly, if we consider that $\|\g\|_1 = m$ (always verified up to a scaling) there exists a value $\rho > \|\g -\vone_m\|_\infty, \rho < 1$ that quantifies the deviation of the gains w.r.t. the ideal case in which they are all equal. Note that $\rho < 1$ also avoids that any component of $\g$ is null, which would correspond to losing all measurements from the corresponding sensor. Hence, the gains will be inside\footnote{$\bb B^m_{\ell_p}$ denotes an $\ell_p$-ball in $\bb R^m$; $\vone^\perp_m$ denotes the orthogonal complement of $\vone_m$, \ie all zero-mean vectors; the projection matrix $\Ponep \coloneqq \I_m - \tfrac{\vone_m \vone^\top_m}{m}$.} $\cl G_\rho \coloneqq \vone_m + \rho {\bb B}^m_{\ell_\infty} \cap \vone^\perp_m$, \ie in a subset $\cl G_\rho \subset \Pi^m_+$. It is in this closed convex set that we search for $\g$. To do so, we start from some $\bs \gamma_0 \in \cl G_\rho$ and compute the projected gradient w.r.t. $\gg$, 
	\begin{equation}
	\bs\nabla^\perp_{\gg} f(\xx,\gg)\coloneqq\tfrac{1}{mp} \sum^p_{l=1}  \Ponep \diag(\A_l \xx) (\diag(\gg) \A_l \xx-\y_l).
	\end{equation}
	This ensures that the steps are taken on $\vone^\perp_m$. In theory, we would have to use the projection operator ${\cl P}_{\cl G_\rho}$ to ensure that a gradient step still belongs to this convex set; however, when we start from an initialisation $\gg_0 \coloneqq \vone_m$, we have observed that the algorithm will remain inside $\cl G_\rho$ when convergent or, conversely, diverge independently of the presence of ${\cl P}_{\cl G_\rho}$. Thus, we will not practically use this projector, while it will be necessary for devising guarantees as in \cite{CambareriJacques2016}.
	
	\begin{figure*}[!t]
		\null\hfill
		\subfloat[{True signal $\x$, $n = \unit[256 \times 256]{px}$}]{
			\includegraphics[width=1.25in]{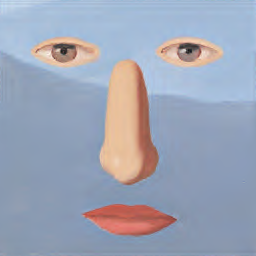}
		}
		\hfill
		\subfloat[{\label{rec:ihtx}Recovery $\hat{\x}$ provided by IHT, ${\rm RSNR}_{\x, \hat \x} =  \unit[17.83]{dB}$}]{
			\includegraphics[width=1.25in]{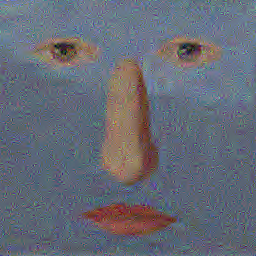}
		}
		\hfill
		\subfloat[{\label{rec:bcihtx}Recovery $\hat{\x}$ provided by BC-IHT, ${\rm RSNR}_{\x, \hat \x} =\unit[153.16]{dB}$}]{
			\includegraphics[width=1.25in]{xh-1.png}
		}
		\hfill\null			
		\\[-1ex]
		\null\hfill
		\subfloat[{True gains $\g$, $m = \unit[103 \times 103]{px}$}]{
			\includegraphics[width=1.9in]{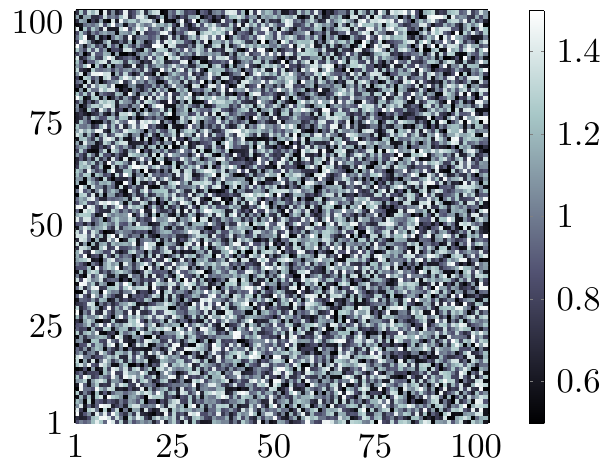}
		}
		\hfill
		\subfloat[{\label{rec:bcihtg}Recovery $\hat{\g}$ provided by BC-IHT, ${\rm RSNR}_{\g, \hat \g} =  \unit[122.76]{dB}$}]{
			\includegraphics[width=1.9in]{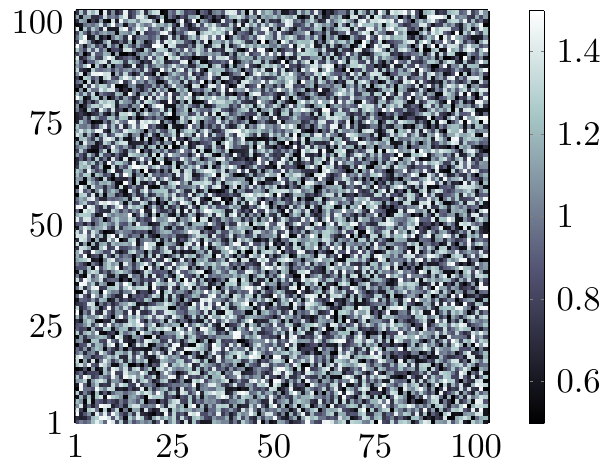}
		}
		\hfill\null			
		\caption{\label{fig:example} {A numerical example of blind calibration for compressive imaging; the test image is a detail of ``Tous les jours'', Ren\'e Magritte, 1966, \copyright ~ Charly Herscovici, with his kind authorization - c/o SABAM-ADAGP, 2011. The artwork was retrieved at {wikiart.org} and is intended for fair use. A comparison of the original and retrieved signal and gains ($\rho = \tfrac{1}{2}$) is reported in a-c and d-e,  respectively.}\vspace{-1.5ex}}
	\end{figure*}
	
	Secondly, as typically done in greedy algorithms, instead of adopting a proxy for sparsity such as the $\ell_1$-norm we iteratively enforce it by evaluating the gradient
	\begin{equation}
	\bs\nabla_{\xx} f(\xx,\gg) = \tfrac{1}{mp} \sum^p_{l=1} \A^\top_l \diag(\gg) (\diag(\gg) \A_l \xx - \y_l)
	\end{equation}
	and applying after each gradient step the {\em hard thresholding} operator ${\cl H}_k$, which sets all but the $k$ largest-magnitude entries of the argument to $0$. This operator is at the heart of Iterative Hard Thresholding (IHT, \cite{BlumensathDavies2009}) and allows us to enforce signal-domain sparsity. Finally, as in \cite{CambareriJacques2016} we choose an initialisation by \emph{backprojection}, \ie $\xx_0\!\coloneqq\!\tfrac{1}{m p} \sum^{p}_{l = 1} \left(\A_l\right)^\top \y_l$ that is an unbiased estimate of $\x$, \ie as $p\rightarrow \infty$ we have that $\xx_0 \rightarrow \bb E \xx_0 \equiv \x$. 
	With all previous considerations, we approach our version of Blind Calibration with Iterative Hard Thresholding (BC-IHT), as summarised in Alg.~\ref{alg1}. The line searches reported in step 3 can be computed in closed form, as they are crucial to accelerate the algorithm (albeit in a sub-optimal fashion). The step-size could be further optimised over the non-linear cost: this may yield faster convergence (see, \eg \cite{Blumensath2012}), but will be the subject of a future improvement of BC-IHT. 
	
	\section{Empirical Phase Transition}
	\label{sec:pht}
	We here propose an extensive experimental assessment of the phase transition of BC-IHT. We explore the effect of the problem dimensions in~\eqref{eq:measurement-model-matrix} on the successful recovery of both the signal and the gains, by varying $n = \{2^9,2^{10}\}$, $k = \{2^5, 2^6, 2^7\}$, $p\!=\!\lceil\{2, 2^{\tfrac{5}{4}}, \ldots,2^5\}\rceil$ and $m\!=\!\lceil \{2, 2^{\tfrac{5}{4}}, \ldots,2^5\} \cdot k\rceil$, while generating $144$ random instances of the problem for each of the configurations. In detail, $\x \sim_{\rm i.i.d.} \cl N^n(0,1)$ is drawn as a standard Gaussian random vector\footnote{The convention $\cl N^{m \times n}(\mu,\sigma^2)$ indicates the generation of an $m\times n$ matrix (or vector) with i.i.d.~Gaussian entries having mean $\mu$ and variance $\sigma^2$.}; $\g$ is drawn uniformly at random on $\cl G_\rho$ for $\rho = \tfrac{1}{2}$; $\A_l \sim_{\rm i.i.d.} \cl N^{m\times n}(0,1)$ are drawn as i.i.d.~Gaussian random matrices. We let the algorithm run given $\y_l$ and $\A_l$, $l\in[p]$ up to a relative change of $10^{-7}$ in the signal and gain updates. Then, we measure the probability of successful recovery
	\[
	{\rm P_\zeta}(n,k,m,p) \coloneqq {\bb P}\left[ \ts\max\left\{\tfrac{\|\hat{\g}-\g\|_2}{\|\g\|_2},\tfrac{\|\hat{\x}-\x\|_2}{\|\x\|_2}\right\}  < \zeta \right]
	\]
	on the trials, with $\zeta = \unit[-60]{dB}$ (this corresponds to an early termination of the algorithm: when convergent, it will reach the exact solution, provided we let it run for a sufficient number of iterations). The results are reported in Fig.~\ref{fig:ept} in terms of the contour levels of $\rm P_\zeta$, as a function of $\log_2 \tfrac{m}{k}$ and $\log_2 p$. 
	
	While a theoretical sample complexity result that grants provable convergence is still under study, we can already appreciate that the effect of increasing $n$ for fixed sparsity levels has a mild effect on the region in which ${\rm P}_{\zeta} > 0.95$, while it does sharpen the transition region as typically observed in standard CS.  Moreover, we can appreciate the impact of increasing $k$ on the transition region while keeping the ratios $\tfrac{m}{k}$ fixed: for larger values of $k$, the region in which the algorithm fails to converge almost surely is rapidly reduced. Moreover, we reported in red the curve that matches $m p = C (k+m)$ (\ie $\log_2 p = \log_2 C(1+\tfrac{k}{m})$) for some $C > 0$, which roughly follows the contours' shape in our experiments.
	
	We highlight that all the empirical evidence collected in this context correctly suggests that $p > 1$: this agrees with our previous finding that $p \gtrsim \log m$ (see \cite[Proposition 2]{CambareriJacques2016}), \ie if no structure is leveraged on the gains $\g$ more than one snapshot will always be needed for the algorithm to collect a sufficient amount of information on $\g$. 
	
	Thus, by interpreting the results, we can expect that if $m \simeq 5 k$ (a widely used rule of thumb in CS), our blind calibration method will converge for most instances of~\eqref{eq:measurement-model-matrix} and $\rho < 1$, once we let $p > 4$ we will be able to recover both $(\x,\g)$. If furthermore $k$ is sufficiently low, the total undersampling factor $\tfrac{mp}{n}$ will be below $1$. 
	
	\section{Blind Calibration for Compressive Imaging}
	\label{sec:img}
	
	We now proceed to apply BC-IHT in a practical case, in which we process a high-dimensional red-green-blue (RGB) image $\x$ of dimension $n = \unit[256\times256]{px}$, which is made sparse w.r.t.~a Daubechies-4 orthonormal wavelet basis with only $k = 1800$ non-zero coefficients. Then $\x$ is acquired by means of Gaussian random sensing matrices $\A_l$, $l \in [p]$. This experiment could be carried out with other sub-Gaussian matrix ensembles such as Bernoulli sensing matrices, with the results being substantially unaltered. 
	Since the sparsity level of the chosen test image is high, we can simulate its acquisition with a sensor array of $m = \unit[103\times103]{px}$ ($m \approx 6 k$) and use $p = 5$ snapshots to meet the requirements of our method; thus $\tfrac{mp}{n} \approx 0.8$, and once the gains are retrieved this CS scheme could revert to $\tfrac{m}{n} \approx 0.16$ while benefiting from the improved model accuracy provided by blind calibration. As for the gains, we set $\rho = \tfrac{1}{2}$ and draw $\g$ uniformly at random from $\cl G_\rho$. 
	
	We then run BC-IHT on each of the RGB channels separately, until the relative change in the signal and gain estimates falls below $10^{-7}$; the quality and data reported below are the worst case among the colour channels. This causes the algorithm to run for $884$ iterations, achieving a high-quality estimate having ${\rm RSNR}_{\bs x, \hat{\bs x}} = -20\log_{10} \tfrac{\|\hat{\x} - \x\|_2}{\|\x\|_2} = \unit[153.16]{dB}$ and ${\rm RSNR}_{\bs g, \hat{\bs g}} = -20\log_{10} \tfrac{\|\hat{\g} - \g\|_2}{\|\g\|_2} = \unit[122.76]{dB}$. The quality of the estimates can be observed in Fig.~\ref{rec:bcihtx} and~\ref{rec:bcihtg}.
	
	To see the beneficial effect of blind calibration, we use the accelerated version of IHT \cite{Blumensath2012} given the exact sparsity level $k$, the snapshots $\y_l$ and the corresponding sensing matrices, which form a standard CS model when concatenated vertically. Hence, accelerated IHT attempts to recover an estimate $\hat{\x}$ while neglecting the model error. The algorithm converges in only $29$ iterations to a local minimiser $\hat{\x}$, whose  ${\rm RSNR}_{\bs x, \hat{\bs x}} = \unit[17.83]{dB}$. Such modest performances can be seen directly in Fig.~\ref{rec:ihtx}. 
	No comparison with other blind calibration algorithms is here explored, since the choice of using a single sparse input and multiple snapshots is specific to our framework. Nevertheless, we note that $(i)$ the computational complexity of our algorithm is competitively low, as it amounts to that of IHT plus an additional projected gradient step in the gain domain per iteration; $(ii)$ just as a proof of convergence for IHT to a local minimiser has been devised, we expect to have provable convergence results in the same fashion, which will lead to a bound on the sample complexity that ensures the retrieval of the exact solution.
	
	\section{Conclusion}
	We proposed a novel approach to blind calibration based on the use of snapshots with multiple draws of the random sensing operator, and on a greedy algorithm which enforces sparsity on the steps resulting from gradient descent on a non-convex objective. Our approach is capable of achieving, within a few snapshots, perfect recovery of the signal and gains in a computationally efficient fashion. Hence, we conclude that when sensor calibration is a cause of concern in a sensing scheme, introducing a modality that follows \eqref{eq:measurement-model-matrix} and using our method could be a viable option to cope with model errors.
	
	We envision that our method may be used both for blind calibration of imaging sensors, as well as distributed sensor arrays or networks if suitably modified to allow for compressive sensing. While we presented empirical evidence on the phase transition of our algorithm, a more rigorous convergence guarantee is the subject of our current study and will be presented in a future communication. 
	
	
	\bibliographystyle{IEEEtran}
	\bibliography{icasspbiblio}

\begin{thebibliography}{10}
\providecommand{\url}[1]{#1}
\csname url@samestyle\endcsname
\providecommand{\newblock}{\relax}
\providecommand{\bibinfo}[2]{#2}
\providecommand{\BIBentrySTDinterwordspacing}{\spaceskip=0pt\relax}
\providecommand{\BIBentryALTinterwordstretchfactor}{4}
\providecommand{\BIBentryALTinterwordspacing}{\spaceskip=\fontdimen2\font plus
\BIBentryALTinterwordstretchfactor\fontdimen3\font minus
  \fontdimen4\font\relax}
\providecommand{\BIBforeignlanguage}[2]{{%
\expandafter\ifx\csname l@#1\endcsname\relax
\typeout{** WARNING: IEEEtran.bst: No hyphenation pattern has been}%
\typeout{** loaded for the language `#1'. Using the pattern for}%
\typeout{** the default language instead.}%
\else
\language=\csname l@#1\endcsname
\fi
#2}}
\providecommand{\BIBdecl}{\relax}
\BIBdecl

\bibitem{Donoho2006}
D.~L. Donoho, ``{C}ompressed sensing,'' \emph{{IEEE} {T}rans. {I}nf. {T}heory},
  vol.~52, no.~4, pp. 1289--1306, 2006.

\bibitem{HermanStrohmer2010}
M.~A. Herman and T.~Strohmer, ``{General deviants: An analysis of perturbations
  in compressed sensing},'' \emph{{IEEE} {J}. {S}el. {T}op. {S}ignal
  {P}rocess.}, vol.~4, no.~2, pp. 342--349, 2010.

\bibitem{Romberg2009}
J.~Romberg, ``{C}ompressive sensing by random convolution,'' \emph{{SIAM} {J}.
  {I}mag. {S}ci.}, vol.~2, no.~4, pp. 1098--1128, 2009.

\bibitem{BjorklundMagli2013}
T.~Bjorklund and E.~Magli, ``{A} parallel compressive imaging architecture for
  one-shot acquisition,'' in \emph{2013 IEEE Picture Coding Symposium (PCS)},
  2013, pp. 65--68.

\bibitem{LiutkusMartinaPopoffEtAl2014}
A.~Liutkus, D.~Martina, S.~Popoff, G.~Chardon, O.~Katz, G.~Lerosey, S.~Gigan,
  L.~Daudet, and I.~Carron, ``{I}maging with nature: {C}ompressive imaging
  using a multiply scattering medium,'' \emph{{S}ci. {R}ep.}, vol.~4, 2014.

\bibitem{BahmaniRomberg2015}
S.~Bahmani and J.~Romberg, ``{C}ompressive deconvolution in random mask
  imaging,'' \emph{{IEEE} {T}rans. {C}omput. {I}maging}, vol.~1, no.~4, pp.
  236--246, 2015.

\bibitem{DumasLodhiBajwaEtAl2016}
J.~P. Dumas, M.~A. Lodhi, W.~U. Bajwa, and M.~C. Pierce, ``{C}omputational
  imaging with a highly parallel image-plane-coded architecture: challenges and
  solutions,'' \emph{{O}pt. {E}xpress}, vol.~24, no.~6, pp. 6145--6155, Mar.
  2016.

\bibitem{HayatTorresArmstrongEtAl1999}
M.~M. Hayat, S.~N. Torres, E.~Armstrong, S.~C. Cain, and B.~Yasuda,
  ``{S}tatistical algorithm for nonuniformity correction in focal-plane
  arrays,'' \emph{{A}ppl. {O}pt.}, vol.~38, no.~5, pp. 772--780, 1999.

\bibitem{CambareriJacques2016}
V.~Cambareri and L.~Jacques, ``{A} non-convex blind calibration method for
  randomised sensing strategies,'' in \emph{2016 4\textsuperscript{th}
  International Workshop on Compressed Sensing Theory and its Applications to
  Radar, Sonar and Remote Sensing (CoSeRa)}, Sep. 2016, pp. 16--20.

\bibitem{CambareriJacques2016a}
------, ``{T}hrough the {H}aze: {A} {N}on-{C}onvex {A}pproach to {B}lind
  {C}alibration for {L}inear {R}andom {S}ensing {M}odels,'' \emph{ar{X}iv
  preprint ar{X}iv:1610.09028}, 2016, submitted to Information and Inference: A
  Journal of the IMA.

\bibitem{Vershynin2012}
R.~Vershynin, ``{I}ntroduction to the non-asymptotic analysis of random
  matrices,'' in \emph{Compressed Sensing: Theory and Applications}.\hskip 1em
  plus 0.5em minus 0.4em\relax Cambridge University Press, 2012, pp. 210--268.

\bibitem{BalzanoNowak2008}
L.~Balzano and R.~Nowak, ``{Blind calibration of networks of sensors: Theory
  and algorithms},'' in \emph{Networked Sensing Information and Control}.\hskip
  1em plus 0.5em minus 0.4em\relax Springer, 2008, pp. 9--37.

\bibitem{LiporBalzano2014}
J.~Lipor and L.~Balzano, ``{R}obust blind calibration via total least
  squares,'' in \emph{2014 IEEE International Conference on Acoustics, Speech
  and Signal Processing (ICASSP)}, 2014, pp. 4244--4248.

\bibitem{RepettiBirdiDabbechEtAl2017}
A.~Repetti, J.~Birdi, A.~Dabbech, and Y.~Wiaux, ``{A} non-convex optimization
  algorithm for joint {DDE} calibration and imaging in radio interferometry,''
  \emph{ar{X}iv preprint ar{X}iv:1701.03689}, 2017.

\bibitem{ZhuLeusGiannakis2011}
H.~Zhu, G.~Leus, and G.~B. Giannakis, ``{S}parsity-cognizant total
  least-squares for perturbed compressive sampling,'' \emph{{IEEE} {T}rans.
  {S}ignal {P}rocess.}, vol.~59, no.~5, pp. 2002--2016, 2011.

\bibitem{ParkerCevherSchniter2011}
J.~Parker, V.~Cevher, and P.~Schniter, ``{C}ompressive sensing under matrix
  uncertainties: {An Approximate Message Passing} approach,'' in \emph{2011
  45\textsuperscript{th} {Asilomar} {Conference} on {Signals}, {Systems} and
  {Computers}}, Nov. 2011, pp. 804--808.

\bibitem{BilenPuyGribonvalEtAl2014}
C.~Bilen, G.~Puy, R.~Gribonval, and L.~Daudet, ``{C}onvex {Optimization
  Approaches} for {Blind Sensor Calibration Using Sparsity},'' \emph{{IEEE}
  {T}rans. {S}ignal {P}rocess.}, vol.~62, no.~18, pp. 4847--4856, Sep. 2014.

\bibitem{FriedlanderStrohmer2014}
B.~Friedlander and T.~Strohmer, ``{B}ilinear compressed sensing for array
  self-calibration,'' in \emph{2014 48\textsuperscript{th} {Asilomar}
  {Conference} on {Signals}, {Systems} and {Computers}}, Nov. 2014, pp.
  363--367.

\bibitem{SchuelkeCaltagironeZdeborova2015}
C.~Sch{\"u}lke, F.~Caltagirone, and L.~Zdeborov{\'a}, ``{B}lind sensor
  calibration using approximate message passing,'' \emph{{J}. {S}tat. {M}ech:
  {T}heory {E}xp.}, vol. 2015, no.~11, p. P11013, 2015.

\bibitem{LingStrohmer2015}
S.~Ling and T.~Strohmer, ``{S}elf-calibration and biconvex compressive
  sensing,'' \emph{{I}nverse {P}rob.}, vol.~31, no.~11, p. 115002, 2015.

\bibitem{AhmedRechtRomberg2014}
A.~Ahmed, B.~Recht, and J.~Romberg, ``{B}lind deconvolution using convex
  programming,'' \emph{{IEEE} {T}rans. {I}nf. {T}heory}, vol.~60, no.~3, pp.
  1711--1732, 2014.

\bibitem{StoegerJungKrahmer2016}
D.~St{\"o}ger, P.~Jung, and F.~Krahmer, ``{B}lind deconvolution and compressed
  sensing,'' in \emph{2016 4\textsuperscript{th} International Workshop on
  Compressed Sensing Theory and its Applications to Radar, Sonar and Remote
  Sensing (CoSeRa)}, 2016, pp. 24--27.

\bibitem{CandesLiSoltanolkotabi2015}
E.~Cand\`es, X.~Li, and M.~Soltanolkotabi, ``{P}hase {Retrieval} via {Wirtinger
  Flow}: {Theory} and {Algorithms},'' \emph{{IEEE} {T}rans. {I}nf. {T}heory},
  vol.~61, no.~4, pp. 1985--2007, Apr. 2015.

\bibitem{SanghaviWardWhite2016}
S.~Sanghavi, R.~Ward, and C.~D. White, ``{T}he local convexity of solving
  systems of quadratic equations,'' \emph{{R}esults {M}ath.}, pp. 1--40, 2016.

\bibitem{SunQuWright2016}
J.~Sun, Q.~Qu, and J.~Wright, ``{A} geometric analysis of phase retrieval,'' in
  \emph{2016 IEEE International Symposium on Information Theory (ISIT)}, Jul.
  2016, pp. 2379--2383.

\bibitem{LiLingStrohmerEtAl2016}
X.~Li, S.~Ling, T.~Strohmer, and K.~Wei, ``{R}apid, {R}obust, and {R}eliable
  {B}lind {D}econvolution via {N}onconvex {O}ptimization,'' \emph{ar{X}iv
  preprint ar{X}iv:1606.04933}, 2016.

\bibitem{MansourKamilov2016}
H.~Mansour and U.~S. Kamilov, ``{M}ultipath removal by online blind
  deconvolution in through-the-wall-imaging,'' in \emph{2016 IEEE International
  Conference on Acoustics, Speech and Signal Processing (ICASSP)}, Mar. 2016,
  pp. 3106--3110.

\bibitem{LeeTianRomberg2016}
K.~Lee, N.~Tian, and J.~Romberg, ``{F}ast and guaranteed blind multichannel
  deconvolution under a bilinear system model,'' \emph{ar{X}iv preprint
  ar{X}iv:1610.06469}, 2016.

\bibitem{LeeLiJungeEtAl2017}
K.~Lee, Y.~Li, M.~Junge, and Y.~Bresler, ``{B}lind {R}ecovery of {S}parse
  {S}ignals {F}rom {S}ubsampled {C}onvolution,'' \emph{{IEEE} {T}rans. {I}nf.
  {T}heory}, vol.~63, no.~2, pp. 802--821, Feb. 2017.

\bibitem{LingStrohmer2016}
S.~Ling and T.~Strohmer, ``{S}elf-{C}alibration via {L}inear {L}east
  {S}quares,'' \emph{ar{X}iv preprint ar{X}iv:1611.04196}, 2016.

\bibitem{ChoudharyMitra2013}
S.~Choudhary and U.~Mitra, ``{O}n identifiability in bilinear inverse
  problems,'' in \emph{2013 IEEE International Conference on Acoustics, Speech
  and Signal Processing (ICASSP)}, 2013, pp. 4325--4329.

\bibitem{ChoudharyMitra2014}
------, ``{S}parse blind deconvolution: {W}hat cannot be done,'' in \emph{2014
  IEEE International Symposium on Information Theory (ISIT)}, Jun. 2014, pp.
  3002--3006.

\bibitem{LiLeeBresler2016}
Y.~Li, K.~Lee, and Y.~Bresler, ``{Identifiability in Blind Deconvolution With
  Subspace or Sparsity Constraints},'' \emph{{IEEE} {T}rans. {I}nf. {T}heory},
  vol.~62, no.~7, pp. 4266--4275, Jul. 2016.

\bibitem{LiLeeBresler2017}
------, ``{I}dentifiability in {B}ilinear {I}nverse {P}roblems {W}ith
  {A}pplications to {S}ubspace or {S}parsity-{C}onstrained {B}lind {G}ain and
  {P}hase {C}alibration,'' \emph{{IEEE} {T}rans. {I}nf. {T}heory}, vol.~63,
  no.~2, pp. 822--842, Feb. 2017.

\bibitem{BlumensathDavies2009}
T.~Blumensath and M.~E. Davies, ``{I}terative hard thresholding for compressed
  sensing,'' \emph{{A}ppl. {C}omput. {H}armon. {A}nal.}, vol.~27, no.~3, pp.
  265--274, 2009.

\bibitem{Blumensath2012}
T.~Blumensath, ``{A}ccelerated iterative hard thresholding,'' \emph{{S}ignal
  {P}rocess.}, vol.~92, no.~3, pp. 752--756, 2012.

\end{thebibliography}
	
\end{document}